\documentclass{PoS}

\def\gs{\mathrel{\hbox{\rlap{\hbox{\lower4pt\hbox{$\sim$}}}\hbox{$>$}}}}
\def\ls{\mathrel{\hbox{\rlap{\hbox{\lower4pt\hbox{$\sim$}}}\hbox{$<$}}}}

\def\suzaku{{\it Suzaku}}
\def\nustar{{\it NuSTAR}}

\def\rosat{{\it ROSAT}}

\def\asca{{\it ASCA}}

\def\swift{{\it Swift}}

\def\xmm{{\it XMM-Newton}}

\def\nustar{{\it NuSTAR}}

\def\et{{et al.\ }}
\def\1h07{{1H~0707--495}}
\def\iras13{{IRAS~13224--3809}}

\def\mrk335{{Mrk~335}}
\def\rej1034{{RE~J1034+396}}
\def\rxj04{{RX~J0439.6-5311}}
\def\izw1{{I~Zw~1}}
\def\rg{{\thinspace r_{\rm g}}}

\def\feka{{Fe~K$\alpha$}}
\def\fela{{Fe~L$\alpha$}}

\def\oiii{{[O~\textsc{iii}]}}

\def\feii{{Fe~\textsc{ii}}}

\def\LLedd{{L/L_{\rm Edd}}}

\def\keV{{\rm\thinspace keV}}

\def\s{{\rm\thinspace s}}
\def\ks{{\rm\thinspace ks}}

\title{X-ray perspective of Narrow-line Seyfert 1 galaxies}

\ShortTitle{X-ray perspective of NLS1s }

\author{\speaker{L. C. Gallo}\\
        Department of Astronomy and Physics, Saint Mary's University, 923 Robie Street, Halifax, NS, B3H 3C3, Canada\\
        E-mail: \email{lgallo@ap.smu.ca}}

\abstract{It is arguably in the X-ray regime that Narrow-line Seyfert 1 galaxies (NLS1s) exhibit the most extreme behaviour.  Spectral complexity, rapid and large amplitude flux variations, and exceptional spectral variability are well known characteristics.  However, NLS1s are not eccentric, but form a continuous sequence with typical Seyfert 1 galaxies.  Understanding the extreme behaviour displayed by NLS1s will provide insight to the general AGN phenomenon.  In this review, I will examine some of the important NLS1 X-ray discoveries over the past twenty years.  I will then explore recent work that looks at the nature of the primary X-ray source (i.e. the corona) in NLS1s, demonstrating how the corona can be compact, dynamic, and in some cases consistent with collimated outflow.  X-ray observations of NLS1s will be key in determining the nature of the corona, resolving the disc-jet connection, and determining the origin of the radio loud/quiet dichotomy in AGN.}

\FullConference{Revisiting narrow-line Seyfert 1 galaxies and their place in the Universe - NLS1 Padova\\
		9-13 April 2018 \\
		Padova Botanical Garden, Italy}

\begin{document}

\section{The X-ray history of NLS1s}

With regards to the general class properties of Seyfert galaxies, Narrow-line Seyfert 1 galaxies (NLS1s) often occupy the extremes of various parameter spaces (e.g. \cite{bg92}, \cite{grupe04a}, \cite{grupe04b}, \cite{komossa18}).  Identified by their optical properties (\cite{osterbrock85}, \cite{goodrich89}), NLS1s exhibit strong \feii\ emission, narrow permitted lines, and weak \oiii.  The properties and behaviour of NLS1s are often attributed to an AGN with a small supermassive black hole and high Eddington ratio ($\LLedd$), perhaps a young system that is aggressively accreting (e.g \cite{grupe96}, \cite{pounds95}).  

Infrared studies find the host-galaxy is being rejuvenated as NLS1s, on average, exhibit higher rates of star formation than typical  broad-line Seyfert 1s (BLS1s) (e.g. \cite{sani10}).  While most NLS1s are radio-quiet, several sources have now been discovered that are radio-loud and some emit gamma-rays (e.g. \cite{foschini12}, \cite{foschini15}, \cite{komossa18}, see section 3 of \cite{komossa18} for summary of the X-ray properties of radio-loud NLS1s). NLS1s display a mix of radio morphologies and many are seen as Compact Steep Spectrum objects (CSS) (\cite{berton18}). The indication of beamed emission in some NLS1s suggests that orientation effects play a part, for at least some sources.

It is arguably in the X-ray regime that NLS1s display the most extreme behaviour.  NLS1s achieved notoriety in the \rosat\ era from the work of Boller, Brandt and Fink (\cite{bbf96}) (see also \cite{grupe96}, \cite{p92}) that demonstrated  these AGN exhibited steeper X-ray spectra between $0.2-2\keV$ than their broad-line counterparts.  The X-ray variability was also remarkable, with objects like \iras13\ (\cite{boller97}) and PHL~1092 (\cite{brandt99}) showing the most rapid and largest amplitude variations seen in radio-quiet objects.  

\asca\ observations confirmed previous known behaviour and expanded the list of radical X-ray properties seen in NLS1s (e.g. \cite{leighly99a}, \cite{leighly99b}, \cite{dewangan02}).  With sensitive up to 10 keV, it was determined that the $2-10$ keV photon index was also steeper for NLS1s than for BLS1s (\cite{brandt97}).  This was perhaps the first sign that the primary X-ray source (i.e. the corona) of NLS1s might be different than in other Seyferts.  The steeper spectrum suggests that the coronae of NLS1s are more diffuse or at a lower temperature than in BLS1s.

\begin{figure}
     \includegraphics[trim= 3cm 2cm 0cm 1cm, width=.52\textwidth]{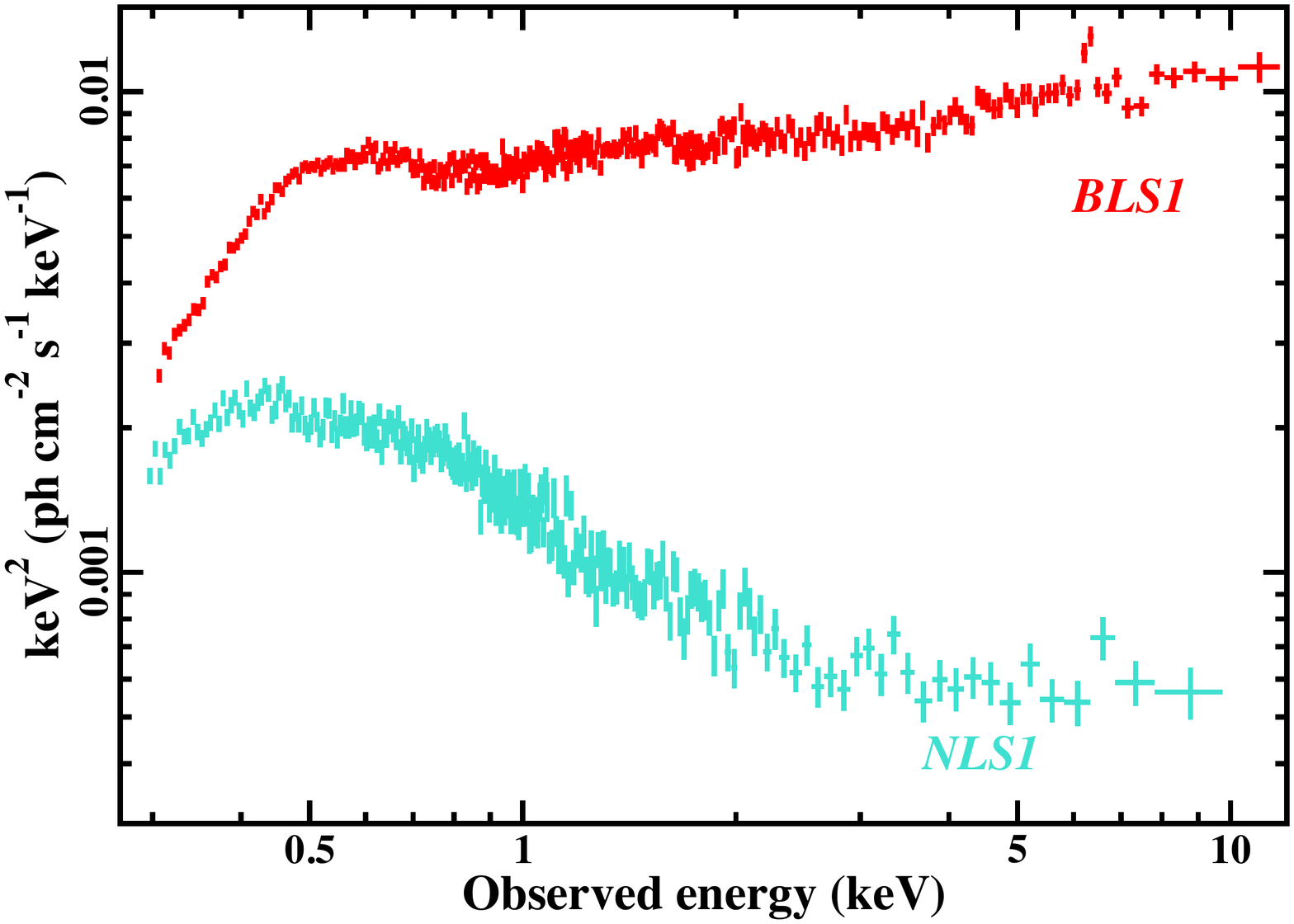}
     \includegraphics[trim= 3cm 2cm 0cm 0cm, width=.52\textwidth]{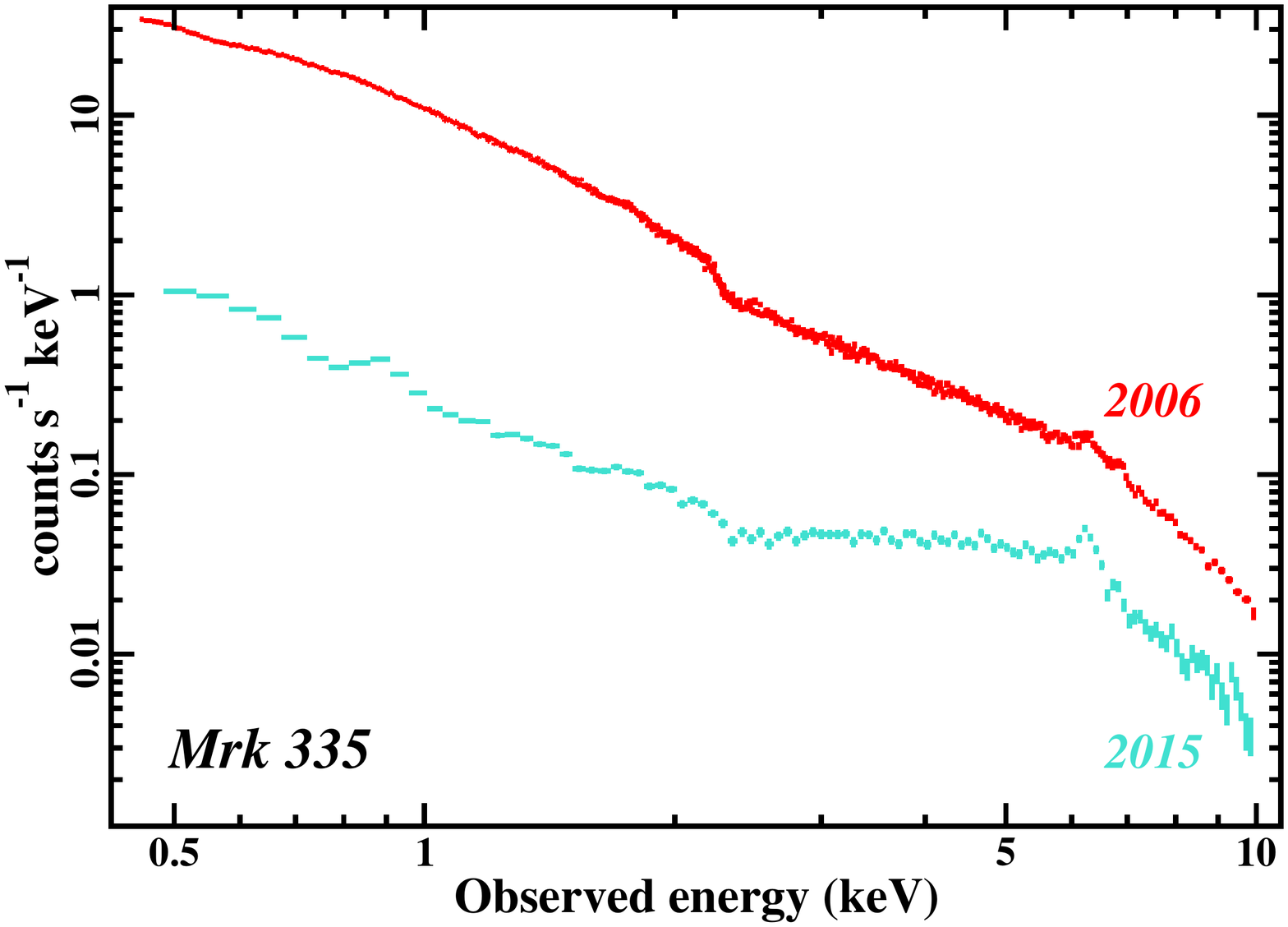}
     \caption{Left:  The $0.3-10\keV$ spectrum of a BLS1 (Mrk~79, red) and a NLS1 (RX J1355+5612, turquoise) compared to a flat power law.  The figure is intended to highlight the overall steeper spectrum and stronger soft excess below $\sim1\keV$ in the NLS1.  Right: The $0.4-10\keV$ spectrum of \mrk335\ in 2006 (red) and 2015 (turquoise) demonstrating the extreme spectral and brightness variations often seen in NLS1s.   }
     \label{recap}
\end{figure}

Early X-ray observations established that, on average, compared to BLS1s, NLS1s were seen to: (i) exhibit steeper spectra in the soft (below $\ls 2\keV$) and hard ($\sim2-10\keV$) bands (e.g. \cite{bbf96}, \cite{brandt97}, \cite{grupe10}); (ii) have a stronger soft excess (e.g. \cite{bbf96}, \cite{g04phl}, \cite{grupe10}); (iii) display more extreme spectral variability (e.g. \cite{grupe07}, \cite{miniutti09}, \cite{komossa17}); and (iv) show larger amplitude and more rapid flux variations (e.g.\cite{boller97}, \cite{brandt99}, \cite{g04iras}, \cite{komossa17}) (Fig.~\ref{recap}).   

\section{NLS1s in the \xmm\ era}

With \xmm\ came the era of high-sensitivity X-ray astronomy and the capability to carry out detailed spectral and timing studies of AGN.  NLS1s immediately fell into the spot light with the discovery of sharp, spectral drops at energies above $\sim7\keV$ in some objects (e.g. \cite{boller02}, \cite{boller03}). 

The feature could be interpreted as a strong iron edge, perhaps in outflow, arising from partial covering of the central region by dense clouds traversing the line-of-sight (e.g. \cite{boller02}, \cite{boller03}, \cite{tanaka04}, \cite{g041h}).  Alternatively, the feature could be the blue wing of a relativistically blurred \feka\ emission line (e.g. \cite{fabian04}), enhanced by light bending effects that preferentially illuminate the inner accretion disc (e.g. \cite{miniutti04}).  Both the partial covering and blurred reflection scenario would render the AGN relatively dim in the X-rays, which is consistent with NLS1s exhibiting larger brightness fluctuations in the X-rays relative to other bands, and often appearing X-ray weak compared to their UV luminosity (e.g. \cite{gallo06}, \cite{ranjan11}, \cite{grupe10}).

Deep observations of NLS1s with \xmm\ proved important to understand the nature of spectral drops.  Observations of \1h07\ revealed excess residuals around $1\keV$ that appeared similar to the residuals in the \feka\ band (\cite{fabian09}).  The $\sim1\keV$ feature was identified as \fela\ that appeared boosted because of high iron abundances in the disc (e.g. \cite{reynolds12}).  The same observation also revealed short negative lags between the continuum dominated energy band and the soft excess showing the continuum to lead the reflection dominated band by $<100\s$ (\cite{fabian09}, \cite{zoghbi10}).  In line with the blurred reflection scenario, these lags are reverberation lags arising from light-travel time between the corona (primary continuum) and inner disc (blurred reflection continuum).  

Together, the spectral features and associated timing properties formed strong evidence that the soft excess, in at least some NLS1s, could be interpreted as the blurred reflection component.  A number of other AGN with similar spectral and timing characteristics to \1h07\ have since been discovered (e.g. \cite{ponti10}, \cite{demarco13}, \cite{kara13}).  Interestingly, most of these AGN turn out to be NLS1s pointing to the importance of this sub-class in understanding the inner-most central engine of active galaxies.

For some NLS1s, the soft excess is so strong that additional considerations are required.  For example, in \iras13, adopting a high density disc (e.g. \cite{jiang18}) or returning radiation (\cite{chiang15}, \cite{cunningham}) may be important to fully describe the soft excess.  In other NLS1s, a type of hybrid disc emission is considered.  For objects with high $\LLedd$ and low black hole mass, like \rej1034\ and \rxj04, colour temperature correction is important and sufficient to displace the thermal disc emission into the X-ray band.  Moreover,  Comptonisation can occur in the low-temperature, optically thick accretion disc, in addition to the corona, thereby enhancing the soft excess emission (e.g. \cite{done12}, \cite{jin17a}).

One of the first and more robust detections of a blue-shifted absorption feature above $\sim7\keV$ was in the NLS1 PG~1211+143 (\cite{pounds03}).  The feature is commonly interpreted as an ultrafast outflow demonstrating that NLS1s provide an important laboratory to examine the connection between the accretion disc and potential high-velocity winds in AGN (e.g. \cite{parker17}).  

Jin et al. (\cite{jin17b}) propose that inclination effects are important and that the difference between NLS1s showing high-energy complexity (``complex'' NLS1s, see \cite{gallo06}), and those that do not (``simple'' NLS1s), is a different view through a clumpy accretion disc wind.  The complex NLS1s, which tend to be X-ray weak (\cite{gallo06}) (e.g. \1h07, PG~1211+143), are supposedly seen through the wind whereas in simple NLS1s (e.g. \rej1034, Ton~S180, \rxj04) the central X-ray emitting region is viewed unobscured. 

Alternatively, the blue-shifted feature could be associated with a layer of the accretion disc itself (\cite{gf11}, \cite{gf13}).  A highly ionised layer above the disc can imprint absorption features on the reflection spectrum emitted from the accretion disc.  The measured energy shifts in the absorption features arise from motion as the material corotates with the disc (\cite{gf11}, \cite{gf13}, Fabian \et submitted).  The model does not invoke any launching mechanism.  This scenario would generate absorption features that are presumably broader than those predicted in a wind, thus a future instrument (e.g. {\it XRISM} and {\it Athena}) with $Hitomi$-like spectral resolution (\cite{tad16}) will be capable of distinguishing these models.
 
\xmm\ has exposed a plethora of remarkable properties and behaviour in NLS1s.  Black hole spin measurements favour (near) maximum spin values for NLS1s (e.g. \cite{walton13}, \cite{fabian09}, \cite{parker14}, \cite{gallo15}, but see \cite{ranjan16}, \cite{bg16}); the first AGN with a QPO detection was the NLS1, \rej1034 (\cite{gierlinski08}); and the first non-linear rms-flux relation in an accreting source was found in the NLS1, \iras13\ (\cite{alston18}).  It is clear that our understanding of AGN physics can only improve with investigations of NLS1s.

\section{The nature of the corona}

The nature of the primary X-ray source, the corona, is an outstanding question in AGN science.  In the last decade, with  improved data quality and availability of good quality spectra above $\sim10\keV$ (e.g. with \suzaku\ and \nustar), it has become possible to investigate corona models and compare with data (e.g. \cite{fabian15}).  

\begin{figure}
     \includegraphics[width=.5\textwidth]{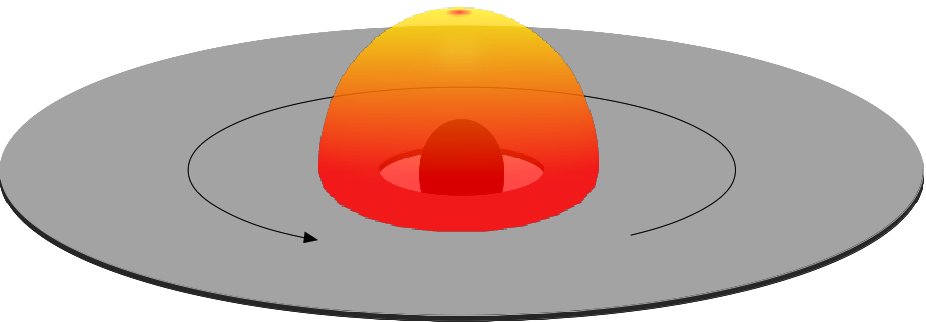}
     \includegraphics[width=.5\textwidth]{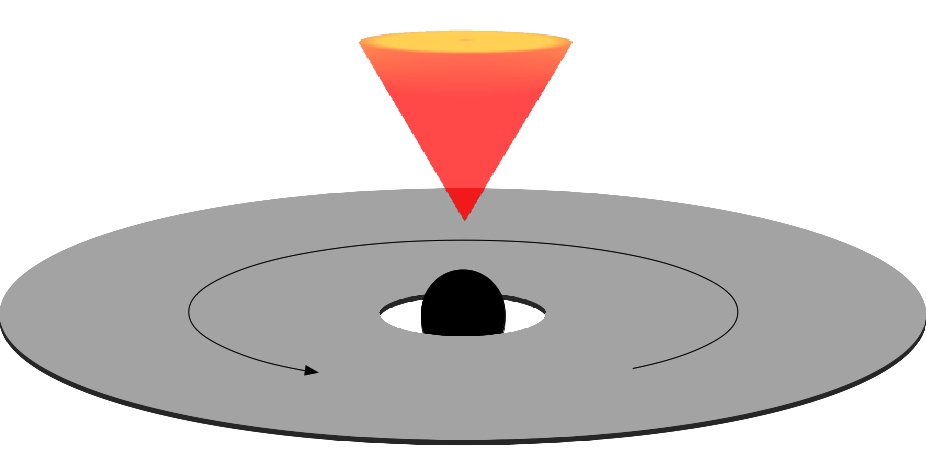}
     \includegraphics[width=.5\textwidth]{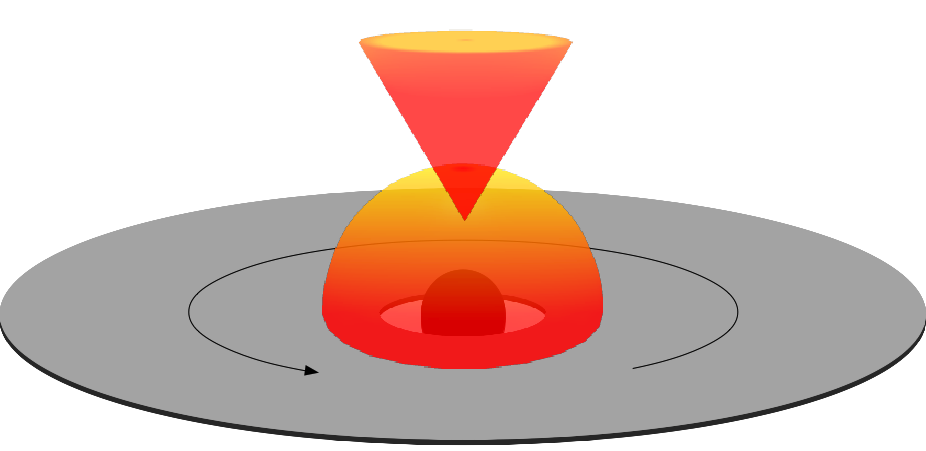}
     \includegraphics[width=.5\textwidth]{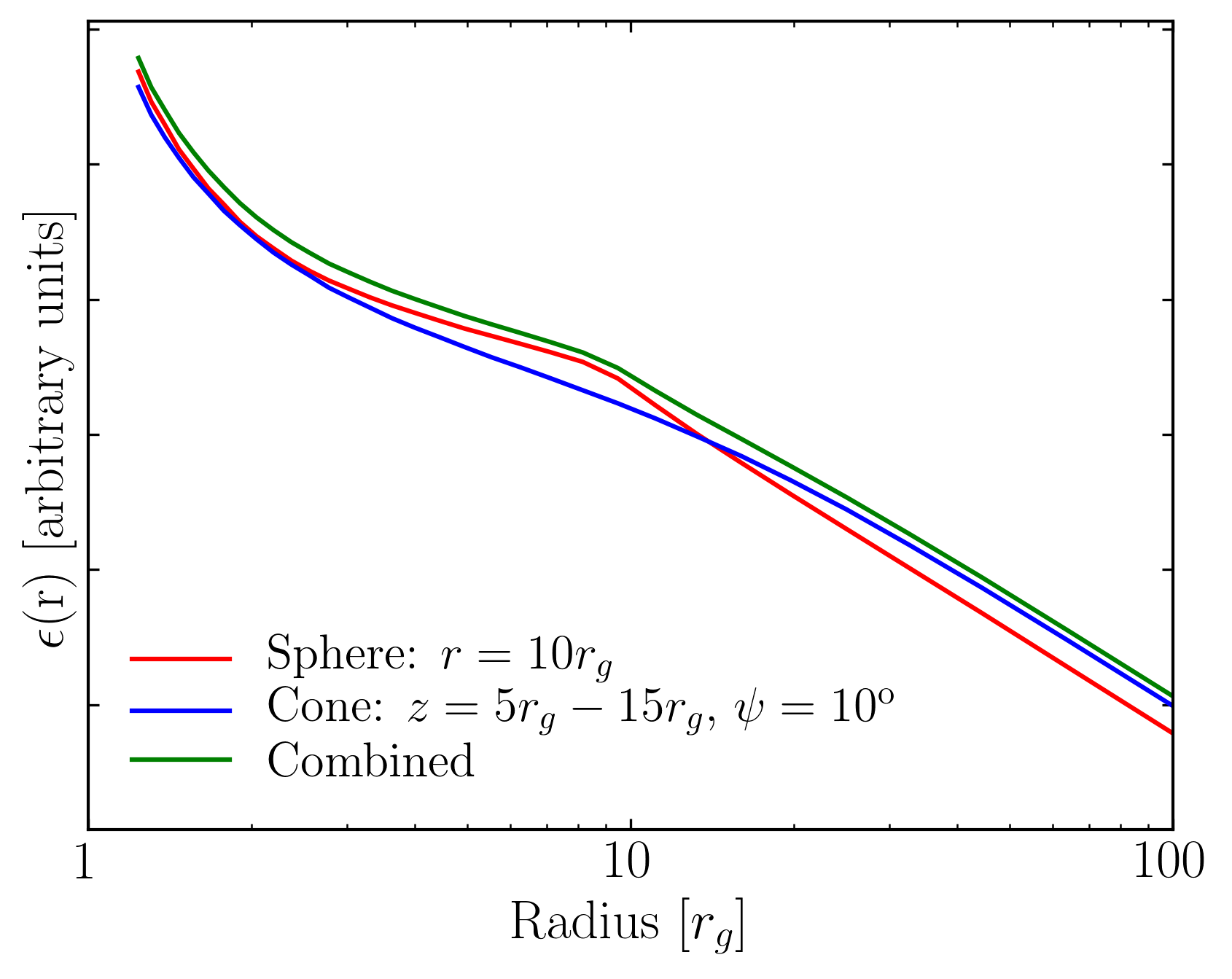}
     \caption{Different coronal geometries like spherical (top left), conical (top right), and blended (lower left) will illuminate the accretion disc differently.  These distinctions can be noted in the emissivity profile (lower right). See \cite{wilkins11}, \cite{wilkins12}, \cite{wg15a}, and \cite{adam17} for details.}
     \label{emis}
\end{figure}

One particularly useful exercise has been to measure the emissivity profile of the \feka\ feature to predict the geometry of the corona (\cite{wilkins11}, \cite{wilkins12}, \cite{wg15a}, \cite{adam17}). To first order, the illumination pattern generated by the corona on to the disc depends on the geometry of the source (Fig.~\ref{emis}).  This information is encoded in the \feka\ emission line profile.   

 \subsection{A breathing corona in \1h07}
 \1h07\ has been observed over a dozen times since the launch of \xmm.  The NLS1 varies significantly in the X-rays and modestly in the UV (e.g. \cite{robertson15}) and always appears X-ray weak (\cite{gallo06}, \cite{ranjan11}).  That is to say the X-rays are dim compared to its UV luminosity (e.g.  \cite{just07}).  
 
 As the X-ray flux varies by about a factor of ten, the spectrum and \feka\ line profile change significantly.  
 Analysis of the emissivity profile in various flux states infers changes in the spatial extent of the X-ray source (\cite{wilkins14}).  The corona apparently contracts when the luminosity is low and expands when it is high.  The average change in radial extent is approximately 30 per cent and indicative of a dynamic corona.
 
Wilkins \et (\cite{wilkins14}) also find that as the source brightens and the corona expands, the spectrum softens.  This implies there are more scattering particles in the larger corona to increase the luminosity, but that the average optical depth and/or the corona temperature decreases to soften the spectrum.
 
Wilkins \et (\cite{wilkins14}) could not address on what timescales the changes in the corona were taking place.  The work collected data over a decade and averaged the observations into various flux states. Observations of a dynamic source that is sufficiently bright to examine such changes as a function of time are required to pin down what is driving the corona to breathe.
 
 \subsection{Collapse and rebuilding of the corona in \mrk335}
In the past decade, \mrk335\ has established itself as an important NLS1 for understanding accretion processes and the nature of the corona in AGN.  Historically, \mrk335\ was one of the brightest AGN in the X-ray sky.  In 2007, the AGN dropped in brightness to approximately one-thirtieth of its typical X-ray flux (\cite{grupe07}, \cite{grupe08}, \cite{longinotti08}).  It has never fully recovered and since 2007 it has remained at about one-tenth its previous X-ray brightness (\cite{grupe12}, \cite{gallo18}).

Continued monitoring with \swift\ (\cite{grupe12}, \cite{gallo18}) shows the AGN is variable in X-rays and UV.  Occasionally, it exhibits deep flux drops (\cite{grupe15}) and spurious flaring events (\cite{grupe14}).  The X-rays can vary by about a factor of 50 and the UV vary by about a factor of 2--3.  The NLS1 is currently in an X-ray weak state (\cite{gallo06}).   The objective of continued monitoring is to understand the cause of this extended X-ray dim state and the origin of the high amplitude flares.

Absorption does play a role in the X-ray observations of \mrk335 (\cite{longinotti13}).  There are indeed multi-phase warm absorbers in modest outflow that leave their mark on the intrinsic continuum spectrum.  It is less clear that the prolonged low-state can be attributed to obscuration of the primary source (e.g. \cite{gallo15}).  Similarly, it is not obvious that there are changes in the nature of the accretion disc.  While \mrk335\ is one of the most variable radio-quiet AGN in the UV (e.g. \cite{grupe12}, \cite{komossa14}), the UV spectrum appears similar in the bright and dim X-ray states (e.g. \cite{gallo18}).  
The changes from the pre-2007 high flux state to the X-ray weak states can be explained as arising from changes in the X-ray corona.  Specifically, during the weak state, the corona has diminished in size and become less luminous (\cite{gallo13},\cite{gallo15}).  

In context of the blurred reflection scenario, the observations of \mrk335\ imply interesting behaviour.  It was recognised rather quickly that the reflection fraction ($\mathcal{R}$, the ratio of the flux in the reflected component and in the intrinsic power law component) was highly variable in \mrk335\ (\cite{gallo13},\cite{gallo15}).  For an isotropic, illuminating source over an infinite plane disc, the reflection fraction should be unity.  If light bending effects are important, than high values of the reflection fraction are measured as coronal emission is directed toward the black hole / inner accretion disc rather than reaching the observer.   The reflection fraction in \mrk335\ varies between very high values, indicative of light bending and reflection dominated emission, to values close to one (e.g. \cite{gallo13},\cite{gallo15}).  

During the 2006 high-flux \suzaku\ observation, when the NLS1 was ten-times brighter than in the 2013 low-flux observation, values of $\mathcal{R}<1$ were measured.  A reflection fraction below one suggests that continuum emission is directed away from the disc, as would occur if the emission were beamed by a source moving away from the disc with notable velocity.  In the 2006 data, the X-ray source could be confined to within $25\rg$ ($25GM/c^2$) if the ejecta did not escape the system.  However, during the 2013 low-flux \suzaku\ observation, the source must be, on average, confined to within $5\rg$ and coronal emission was predominately illuminating the inner accretion disc resulting in a high reflection fraction (\cite{gallo15}).  

The comparison between the high- and low-flux states suggests that the changes could be attributed to a corona that has collapsed.  Changes in the accretion flow or magnetic field structure could be driving the collapse, perhaps similar to what occurs in stellar-mass black hole systems.

\begin{figure}
          \includegraphics[width=1\textwidth]{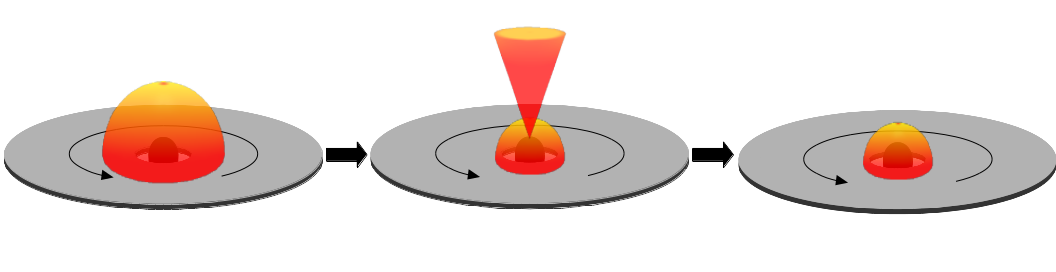}
     \includegraphics[trim= 2cm 1.5cm 1cm 1cm, width=1\textwidth]{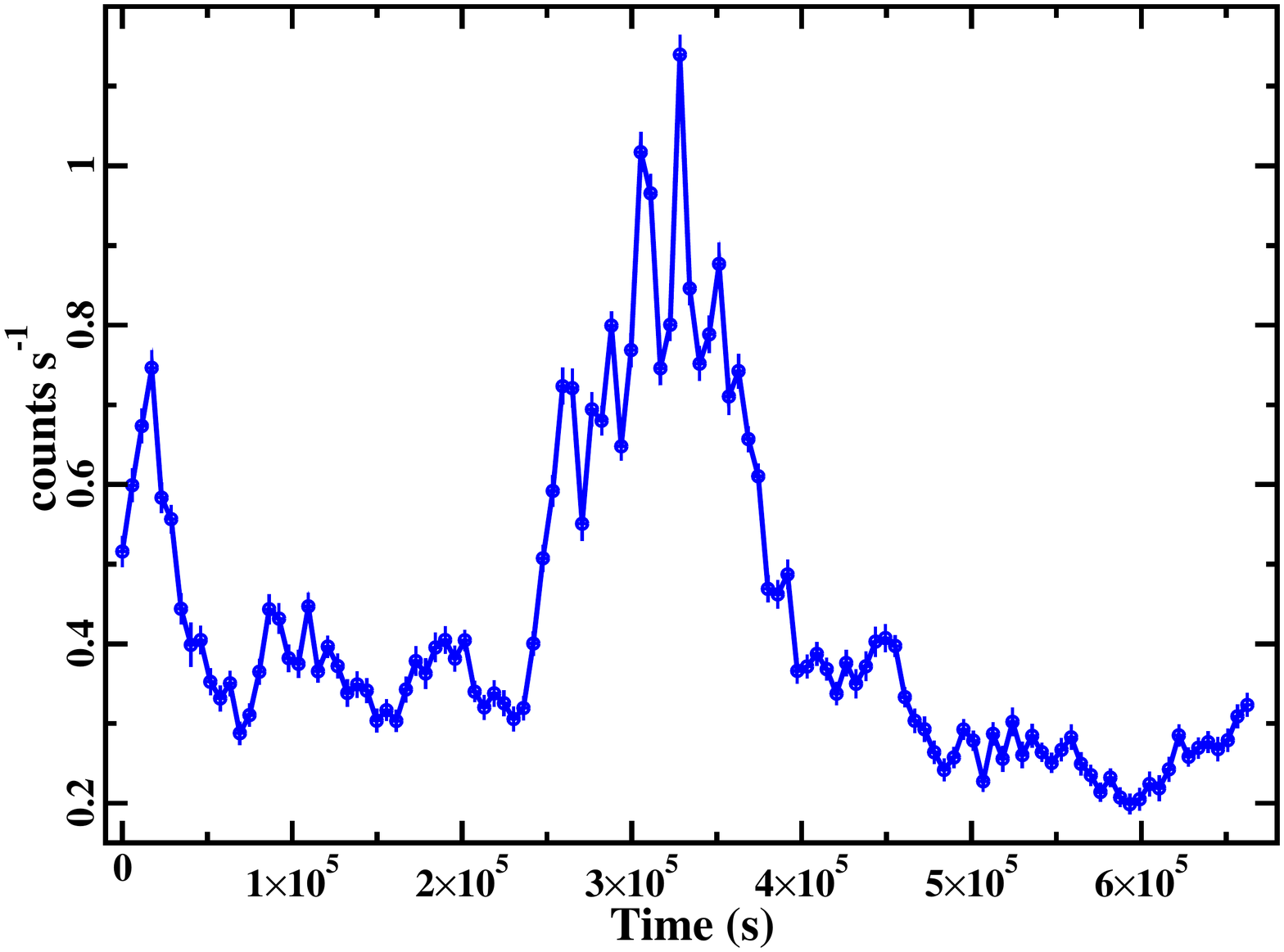}
     \caption{The upper panel provides a depiction of how the corona structure may be changing in \mrk335\ over the duration of the week-long observation (\cite{wg15b}).  The corresponding variations in the light curve are seen in the lower panel.   }
     \label{launch}
\end{figure}

Of interest was the behaviour during a modest X-ray flare in the week-long, low-flux, 2013 \suzaku\ observation.  Over the duration of the flare, there were changes in the reflection fraction, spectral shape, and emissivity profile.  Prior to the flare, the reflection fraction was high ($\mathcal{R}\approx10$), the spectrum hard, and the corona extended out to $\sim7\rg$.  During the flare, the spectrum softened significantly while the reflection fraction dropped to $\mathcal{R}\approx1$ (\cite{wg15b}).  After the flare, the spectrum hardened again, the reflection fraction climbed ($\mathcal{R}>10$), but the extent of the corona seemed to diminish ($\sim3.5\rg$) compared to the pre-flare value.  The measurements infer that during the flare the corona collimated vertically in a jet-like configuration resulting in softening of the spectrum and lowering of the reflection fraction.  As the flare diminished, the jet-like structure dissipated leaving a more compact spherical corona in its place (Fig.~\ref{launch}) (\cite{wg15b}).  Furthermore, this analysis was the first attempt at probing the timescales on which the geometry of the corona could be changing.

The flaring activity in \mrk335\ could be the AGN attempting to launch a jet.  Like most NLS1s, \mrk335\ is radio-quiet, but there could be a small scale structure like the base of an aborted jet present \cite{ghisellini03}.  This feature could be common in radio-quiet AGN \cite{ghisellini03} and could serve as a reservoir of energetic particles for the extended spherical corona. 

The circumstantial evidence of rapid ejection from the coronal region was corroborated during a giant amplitude 2014 X-ray flare (\cite{grupe14}) that triggered a \nustar\ TOO (PI Gallo).  The observation showed a reflection fraction of $\mathcal{R}\approx0.5$ indicating high-velocity beamed emission and a compact source within $5\rg$.  Coincident UV brightening was also recorded during the flare suggesting the brightening in the two bands could be originating from the same process, perhaps synchrotron emission attributed to a jet-like structure (\cite{wilkins15}).

 \subsection{Two stable coronae in \izw1}

\izw1, often called the ``prototypical'' NLS1 because of its strong, optical \feii\ emission, exhibits seemingly atypical properties in the X-rays.  Variable warm absorbers imprint significantly on the low-energy X-ray spectrum (\cite{costantini07},  \cite{silva18}) often rendering studies of the relatively weak soft-excess challenging (\cite{leighly99a}, \cite{g04izw1}).  However, the NLS1 exhibits a strong and variable \feka\ emission line and a steep, power law spectrum (\cite{gallo07a}, \cite{gallo07b}).  A short \xmm\ observation in 2000 revealed X-ray flaring isolated in the $3-10\keV$ band and attributed to coronal flaring (\cite{g04izw1}).

Deeper observations in 2005 revealed further complexity, as the AGN appeared to exhibit distinctly different behaviour from one part of the observation to the next.   The AGN went from displaying distinct changes in spectral shape to a relatively constant spectral shape that varied only in brightness  (\cite{gallo07a}, \cite{gallo07b}).  Interestingly, the transition from one behaviour to the other occurred within kiloseconds and was marked by a sharp dip in the light curve.  Gallo \et (\cite{gallo07a}, \cite{gallo07b}) suggested the behaviour could be described as originating from a two component corona.  One component was radially extended and illuminating more distant parts of the accretion disc, while a more compact corona, perhaps the base of a failed jet (\cite{ghisellini03}),  illuminated the inner few gravitational radii of the disc and was more variable.

The two-coronae scenario was substantiated with a much longer, $270\ks$ \xmm\ observation of \izw1\ in 2015.  The long observation enabled robust timing analysis in addition to spectral modelling (\cite{wilkins17}).  A principal component analysis (PCA) revealed the shape of three variable components.   Two of the eigenvectors (variable components) appeared as power laws  that varied independent of each other (\cite{wilkins17}).  Each of these principal components could be attributed to a different source of coronal emission.

The timing analysis also revealed two high-frequency lags resembling reverberation lags.  The different delay of each lag (i.e. $\sim160\s$ and $60\s$) was interpreted as light-travel time from different parts of the corona.  The lag-energy spectrum could be used to discern the geometry of the illuminating source suggesting that one lag was produced in an extended geometry and the second in a collimated core at the centre (see lower left panel of Fig.~\ref{emis}).

\begin{figure}
     \includegraphics[width=.5\textwidth]{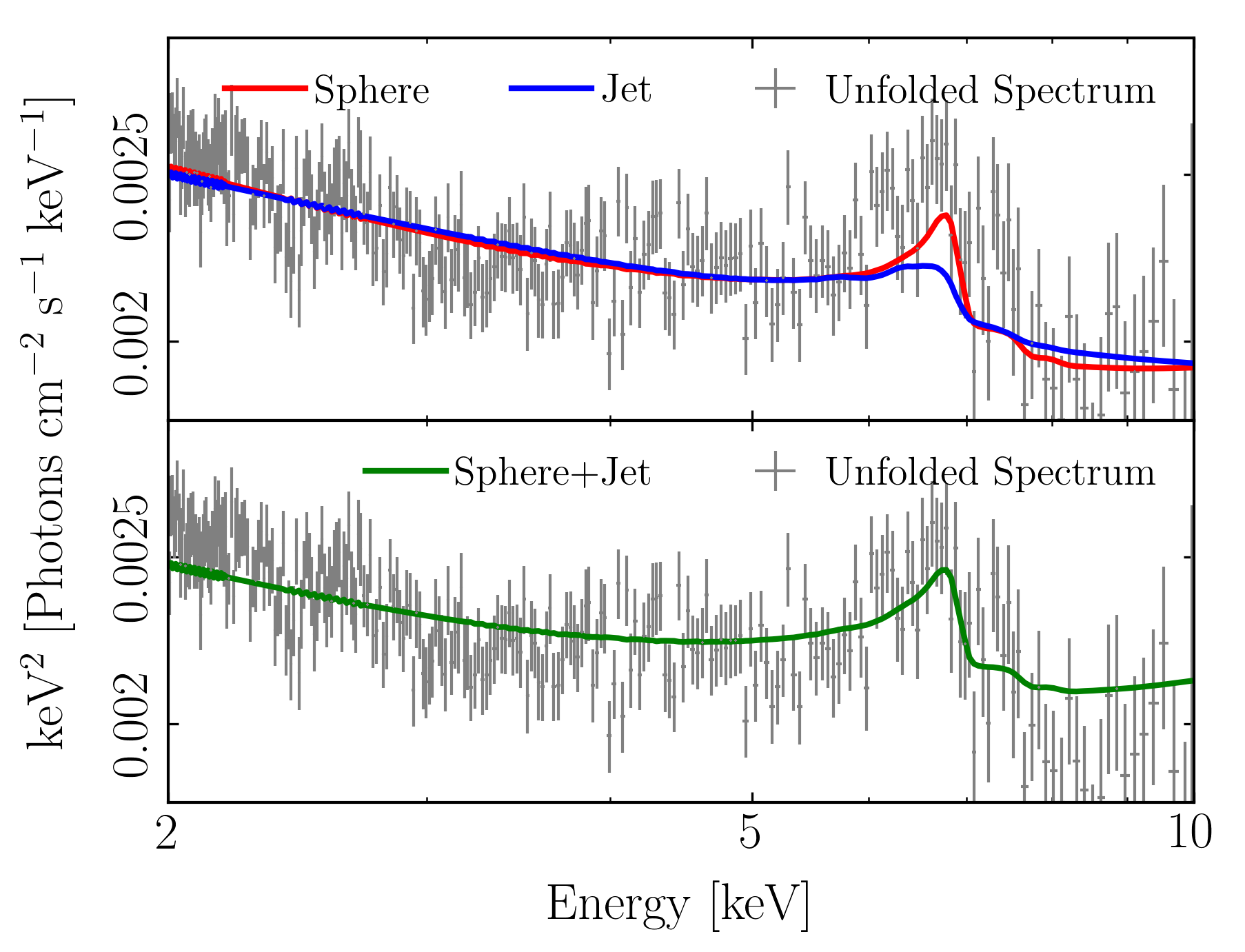}
     \includegraphics[width=.5\textwidth]{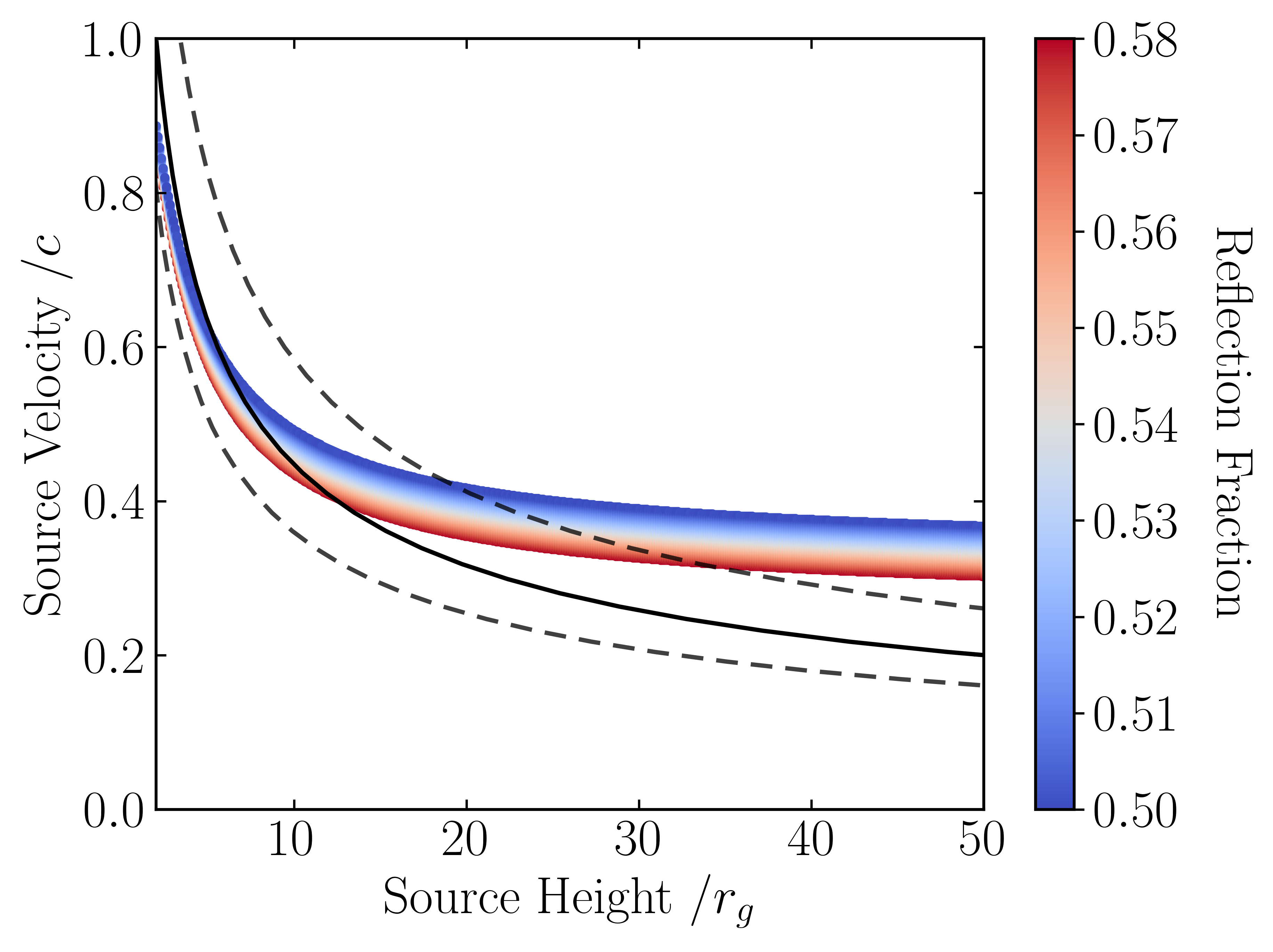}
     \caption{Upper Left:  The disc emissivity profile generated from a corona with a spherical and collimated (jet) geometry is compared to the spectrum of \izw1.  Lower Left:  The illumination pattern produced from a blended corona better represents the data.  Right:  Based on the measured reflection fraction, the velocity of the ejecta in the collimated corona as a function of height can be estimated.  Compared to the escape velocity assuming the estimated black hole mass for \izw1\ (solid and dashed curves) indicates the ejecta could escape the central region. }
     \label{jcemis}
\end{figure}

Spectroscopic modelling of the \xmm\ data corroborate the multiple coronae interpretation (Gallo \et in prep).  In Fig.~\ref{jcemis}, the \feka\ band is shown fitted with a model of a corona that is purely spherical and one that is purely collimated (jet-like) (see Fig.~\ref{emis} for a cartoon depiction of these scenarios).  Neither model adequately fits the \feka\ emission line (Fig.~\ref{jcemis}, upper left panel).  However, a combined jet/sphere corona can describe the observed line profile (Fig.~\ref{jcemis}, lower left panel).  Furthermore, with the reflection fraction measured from the spectrum and from equation 15 of Gonzalez et al. (\cite{adam17}), the source velocity can be estimated (Fig.~\ref{jcemis}, right panel).  Depending on the initial height of the source (jet) above the black hole, the ejecta could exceed the escape velocity in \izw1\ (Gallo \et in prep).

\section{Putting it together}

Almost independent of the parameter space we consider for Seyfert~1 activity, NLS1s typically occupy some extreme of a continuous sequence.  On average, NLS1s exhibit softer X-ray spectra than BLS1s, but significant spectral variability can substantially complicate their appearance.  Most NLS1s that exhibit complexity in their X-ray spectra are X-ray weak (\cite{gallo06}), possessing lower X-ray emission than expected for their measured UV luminosity.  The nature of the X-ray corona may be one of the primary difference between NLS1s and BLS1s.  Recent observations suggest the corona in NLS1s may be more dynamic and less stable than in BLS1s.  Understanding what is driving this behaviour is more difficult. 

Observations of some NLS1s, like \1h07, \mrk335\ and \izw1, provide insight to the coronal processes.  There are indications that the corona is dynamic and its structure (e.g. size) can change on fairly short timescales (e.g. weeks-to-years).  There are even signs that the entire corona could collapse rendering the AGN X-ray weak.  There is evidence that in some NLS1s, at least part of the corona is consistent with a collimated core that is moving away from the disc.  This could be the base of an aborted jet that has been proposed for radio-quiet AGN, and could serve to fill the corona of high-energy particles needed for Comptonisation of disc photons.  Much work is still needed to fully understand the complex behaviour we are observing.

It may be that NLS1s are not a homogeneous group.  Strict adherence to the class definition may exclude some true NLS1s or include impostors. One explanation may not describe the entire class.  Nonetheless, NLS1s clearly have an important role in understanding the AGN phenomenon and for investigating the most compelling questions in AGN science.  X-ray observations of NLS1s will be key in determining the nature of the corona, resolving the disc-jet connection, and determining the origin of the radio loud/quiet dichotomy in AGN.

\section*{Acknowledgements}

Many thanks to the SOC for their invitation to an excellent meeting.
I would like to thank all my collaborators that have worked on NLS1s with me over the years, including Dan Wilkins, Elisa Costantini, Dirk Grupe, Andy Fabian, and Stefanie Komossa.  A special thanks to Adam Gonzalez for producing some figures.  Finally, I would like to dedicate this work to the memory of my mentor and friend Professor Yasuo Tanaka.

This conference has been organized with the support of the
Department of Physics and Astronomy ``Galileo Galilei'', the 
University of Padova, the National Institute of Astrophysics 
INAF, the Padova Planetarium, and the RadioNet consortium. 
RadioNet has received funding from the European Union's
Horizon 2020 research and innovation programme under 
grant agreement No~730562.

\end{document}